\documentclass[aps,prb,twocolumn,amsmath,amssymb,showpacs,superscriptaddress]{revtex4-1}

\usepackage{graphicx}
\usepackage{amsmath,amssymb,amsfonts}
\usepackage{textcomp}
\usepackage{hyperref}
\usepackage{gensymb}
\usepackage{verbatim}
\usepackage{amsmath}
\hypersetup{breaklinks=true,colorlinks=true,urlcolor=black}
\usepackage{color}
\usepackage{soul}
\DeclareGraphicsExtensions{.jpg,.pdf,.png,.eps}

\begin{document}
\title{Avoided ferromagnetic quantum critical point in pressurized La$_5$Co$_2$Ge$_3$}
\author{Li Xiang}
\email[]{ives@iastate.edu}
\author{Elena Gati}
\author{Sergey L. Bud'ko}
\author{Scott M. Saunders}
\author{Paul C. Canfield}
\email[]{canfield@ameslab.gov}
\affiliation{Ames Laboratory, Iowa State University, Ames, Iowa 50011, USA}
\affiliation{Department of Physics and Astronomy, Iowa State University, Ames, Iowa 50011, USA}

\date{\today}

\begin{abstract}
	We present the pressure-temperature phase diagram La$_5$Co$_2$Ge$_3$ up to $\sim$ 5\,GPa, which was constructed from magnetization, resistivity and specific heat measurements. At ambient pressure, La$_5$Co$_2$Ge$_3$ is an itinerant ferromagnet with a Curie temperature $T_\textrm C\sim$ 4\,K. Upon increasing pressure up to $\sim$ 1.7\,GPa, $T_\textrm C$ is suppressed down to $\sim$ 3\,K. Upon further increasing pressure, our results suggest that La$_5$Co$_2$Ge$_3$ enters a different low-temperature ground state. The corresponding transition temperature, $T^*$, has a nonmonotonic pressure dependence up to $\sim$ 5\,GPa. Our results demonstrate that the ferromagnetic quantum critical point in La$_5$Co$_2$Ge$_3$ is avoided by the appearance of a different, likely magnetically ordered state that has an antiferromagnetic component.
\end{abstract}

\maketitle 

\section{Introduction}
Suppressing a second-order phase transition to zero temperature has been of great interest, since exotic physical phenomena, such as unconventional superconductivity, heavy Fermi-liquid etc., are often found in the proximity of the quantum critical point (QCP)\cite{Steglich1979,Dagotto1994,Pfleiderer2001,Paglione2010,Canfield2016}. Whereas antiferromagnetic (AFM) transitions in many metals can be continuously suppressed to zero temperature by a non-thermal tuning parameter, such as pressure, chemical substitution or magnetic field\cite{Gegenwart2008,Shibauchi2014}, striking differences are observed when suppressing ferromagnetic (FM) transitions in metals. Current theoretical models suggest that, when tuning a second-order FM transition in metals towards zero temperature, the quantum criticality is avoided for general reasons. Possible predicted outcomes in clean metallic systems include that, when tuning a second-order FM transition towards zero temperature, the FM transition either becomes of first-order through a tricritical point, or a long-wavelength AFM phase appears\cite{Belitz1999,Chubukov2004,Conduit2009,Karahasanovic2012,Pedder2013,Brando2016RMP}. Whereas a first-order FM transition was experimentally verified in several metallic systems\cite{Huxley2000,Pfleiderer2002,Uhlarz2004,Niklowitz2005,Brando2016RMP}, a modulated magnetic phase was only observed in a few compounds\cite{Kotegawa2013,Cheng2015,Brando2016RMP,Niklowitz2019}. In contrast, it was found that in disordered systems the FM transition remains continuous to low temperatures\cite{Brando2016RMP}. Furthermore, a recent theoretical work proposes that a FM QCP can be realized even in a clean system, when the systems is noncentrosymmetric with a strong spin-orbit interaction\cite{Kirkpatrick2020}. The multiplicity of possible scenarios in itinerant ferromagnets motivated our search for new metallic ferromagnets, in which the (avoided) ferromagnetic criticality is experimentally accessible by using a tuning parameter, which does not introduce any additional disorder, such as hydrostatic pressure.

As part of an ongoing search for fragile magnetic ordering\cite{Canfield2016}, we recently discovered a new itinerant, ferromagnetic compound La$_5$Co$_2$Ge$_3$\cite{Saunders2020}. La$_5$Co$_2$Ge$_3$ belongs to the $R_5$Co$_2$Ge$_3$ ($R$ = La - Sm) family which crystallizes in a monoclinic structure ($C$2/$m$ space group)\cite{Lin2017}. At ambient pressure, thermodynamic, transport, and moun spin relaxation ($\mu$SR) measurements showed that La$_5$Co$_2$Ge$_3$ undergoes a FM transition at $T_\textrm C \simeq$ 3.8\,K. In addition, the magnetism associated with La$_5$Co$_2$Ge$_3$ was found to be itinerant with a low-field saturated moment of $\sim$ 0.1\,$\mu_\textrm B$/Co. These properties make La$_5$Co$_2$Ge$_3$ a rare, small moment, low $T_\textrm C$ compound, which is a promising candidate material for tuning the FM transition towards even lower temperatures.

Motivated by this discovery, in this work we investigate the pressure-temperature phase diagram of La$_5$Co$_2$Ge$_3$ up to 5.12 GPa. To this end, magnetization, resistivity as well as specific heat measurements were performed under pressure. Our study demonstrates that $T_\textrm C$ is suppressed from $\sim$ 4\,K to $\sim$ 3\,K upon increasing pressure up to $\sim$ 1.7\,GPa. Upon further increasing pressure, different resistive and specific heat features are observed. Our results suggest that La$_5$Co$_2$Ge$_3$ enters a different, likely magnetic, low-temperature ground state that has an antiferromagnetic component. Therefore, La$_5$Co$_2$Ge$_3$ is another example, in which ferromagnetic criticality in metals is avoided by the occurrence of a new phase.

\section{Experimental details}
Single crystals of La$_5$Co$_2$Ge$_3$ were grown using a flux method as described in Ref. \onlinecite{Saunders2020}. Low-field (25\,Oe) dc magnetization measurements on a crystal (with magnetic field applied along a random orientation) under pressure were performed in a Quantum Design Magnetic Property Measurement System (MPMS-3) SQUID magnetometer. The measurements were performed on warming after zero-field-cooling from above the magnetic and superconducting transitions of La$_5$Co$_2$Ge$_3$ and Pb manometer respectively. A commercially-available HDM Be-Cu piston-cylinder pressure cell\cite{HDM} was used to apply pressures up to $\sim$ 1\,GPa. Daphne oil 7373, which solidifies at $\sim$ 2.2\,GPa at room temperature\cite{Yokogawa2007}, was used as a pressure medium, ensuring hydrostatic conditions during the pressure change (see below for details). The superconducting transition temperature of elemental Pb was used as a low-temperature manometer\cite{Eiling1981}.

The resistivity measurements with current applied along the crystallographic $b$ ($j\parallel b$) and $c$ ($j\parallel c$) directions were performed in a Quantum Design Physical Property Measurement System (PPMS) using a 1 mA excitation with frequency of 17\,Hz, on cooling using a rate of -0.25\,K/min. A standard, linear four-terminal configuration was used. The magnetic field was always applied perpendicular to the $bc$ plane (i.e., along the $a^*$ direction), along which direction the largest saturated magnetization was observed at ambient pressure\cite{Saunders2020}. To apply pressures up to $\sim$ 2.3\,GPa, a Be-Cu/Ni-Cr-Al hybrid piston-cylinder cell (abbreviated as PCC), similar to the one described in Ref. \onlinecite{Budko1984}, was used. A 4:6 mixture of light mineral oil:n-pentane, which solidifies, at room temperature, in the range $3-4$ GPa\cite{Budko1984,Kim2011,Torikachvili2015}, was used as pressure medium. To apply higher pressures, up to $\sim$ 5.1\,GPa, a modified Bridgman Anvil Cell (mBAC)\cite{Colombier2007} was used. A 1:1 mixture of iso-pentane:n-pentane, which solidifies at $\sim$ 6.5\,GPa at room temperature\cite{Torikachvili2015} was used as the pressure medium for the mBAC. For both types of pressure cells, pressure values at low temperature were inferred from the $T_\textrm{c}(p)$ of lead\cite{Bireckoven1988,Xiang2020}.

Specific heat measurements under pressure up to $\sim$ 2.4\,GPa were performed using an AC calorimetry technique in a PPMS. Details of the setup used and the measurements protocol are described in Ref. \onlinecite{Gati2019}. The same PCC, with same pressure medium and low-temperature pressure gauge, as in resistivity measurements was used.

For all measurements under pressure, the pressure was changed at room temperature and locked by tightening a lock-nut. The pressure variation across the Pb manometer at low temperature can be estimated from the increase of the superconducting transition width with pressure, that can be as large as 0.06\,GPa depending on the cell and absolute pressure. Specifically, for the HDM cell, the PCC and the mBAC with maximum pressures up to $\sim$ 1\,GPa, $\sim$ 2.3\,GPa and $\sim$ 5.1\,GPa, the pressure variations are up to $\sim$ 0.01\,GPa, $\sim$ 0.01\,GPa and $\sim$ 0.06\,GPa, respectively. The measurement results shown and discussed in the main text are taken upon increasing pressure. Data taken upon decreasing pressure are shown and discussed in the Appendix.

\section{Results and discussions}
Figure\,\ref{fig1_MT} shows the temperature-dependent magnetization, $M(T)$, under pressures up to 0.99 GPa. The sharp onset of the diamagnetism at $\sim$ 7\,K is associated with the superconducting transition of elemental Pb, which was used to determine the low-temperature pressure. With decreasing temperature, a rapid increase of the magnetization is observed at $\sim$ 4\,K for all pressures, which is associated with a FM ordering. The transition temperature, $T_\textrm C$, is determined from the intersection of the two dashed lines as indicated in Fig.\,\ref{fig1_MT}. The dashed line on the low-temperature side corresponds to a line, which goes through the point of maximum slope of $M(T)$ and whose slope corresponds to this maximum slope. The dashed line on the high-temperature side is a linear fit to the $M(T)$ data in a 1\,K-temperature window below the Pb $T_\textrm c$ and above the sharp increase of $M$. In order to estimate the uncertainty of our $T_\textrm C$ determination, we have used multiple 1\,K windows in this limited temperature range. $T_\textrm C$ is suppressed from $\sim$ 4\,K to $\sim$ 3.8\,K upon increasing pressure from 0.16\,GPa to 0.99\,GPa. Finally, the decrease of $M$ below $\sim$ 3\,K, observed in low-field magnetization measurements after zero-field cooling, could be related to the formation of ferromagnetic domains in the crystal.


\begin{figure}
	\includegraphics[width=8.6cm]{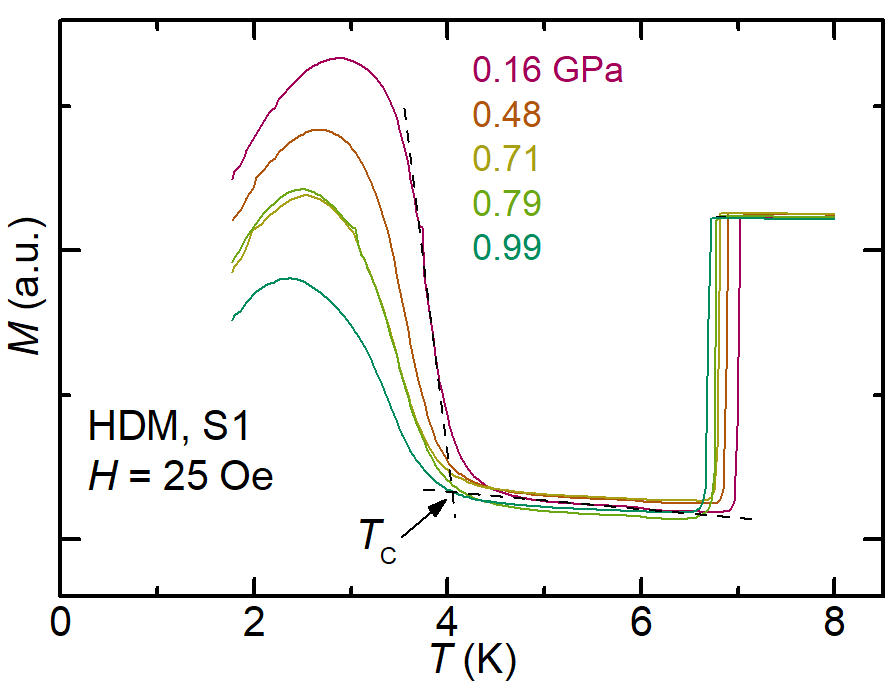}%
	\caption{Temperature dependence of the magnetization, $M(T)$, after zero-field cooling (ZFC) of La$_5$Co$_2$Ge$_3$ (sample S1) under hydrostatic pressures up to 0.99\,GPa in a HDM Be-Cu piston-cylinder pressure cell with an applied field of 25 Oe. The superconducting transition of elemental Pb, which gives rise to the sharp drop of the magnetization at $T\sim$ 7\,K, is used to determine the low-temperature pressure values. The criterion for the determination of the ferromagnetic transition $T_\textrm C$ is indicated by the dashed lines and arrow in the figure (see text for details).
		\label{fig1_MT}}
\end{figure}

\begin{figure}
	\includegraphics[width=\linewidth]{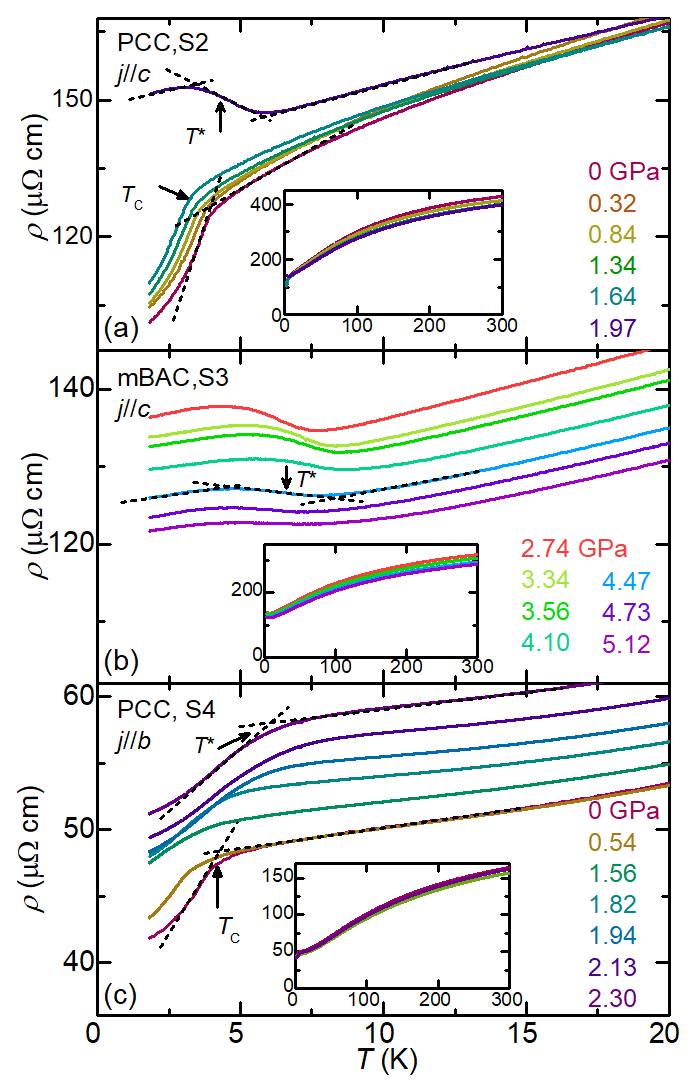}%
	\caption{Resistivity measurements under pressure on La$_5$Co$_2$Ge$_3$. (a-c) Low-temperature resistivity, $\rho(T)$, for sample S2 measured in a piston-cylinder cell with current applied along $c$ (a), for sample S3 measured in a modified Bridgman Anvil cell with current applied along $c$ (b), and for sample S4 measured in a piston-cylinder cell with current applied along $b$ (c). Insets: $\rho(T)$ curves in the full temperature range up to 300\,K. Criteria for the determination of the ferromagnetic transition temperature $T_\textrm C$ and the transition temperature into new ground state $T^*$ are indicated by dashed lines and arrows in the figures (see text for details). Data curves in the main panels of (a) and (c) are shifted up by 2\,$\mu\Omega$ cm for clarity.
		\label{fig2_RT}}
\end{figure}

To investigate the phase diagram to higher pressure, resistivity measurements on several specimens were performed utilizing different pressure cells. Specifically, samples S2, S3 and S4 were measured in the PCC, the mBAC and the PCC with $j\parallel c$, $j\parallel c$ and $j\parallel b$, respectively. The results are summarized and presented in Fig. \ref{fig2_RT}. At ambient pressure, in agreement with Ref. \onlinecite{Saunders2020}, for resistivity measured with $j\parallel b$ and $j\parallel c$, a sharp drop of resistivity is observed at $T\sim$ 4\,K which is associated with the FM transition. In addition, the $c$-axis resistivity shows a downturn curvature (d$^2\rho$/d$T^2<$ 0) for $T>T_\textrm C$ (see Fig. \ref{fig2_RT} (a) inset), whereas the $b$-axis resistivity shows a upturn curvature (d$^2\rho$/d$T^2>$ 0) for $T_\textrm C<T\lesssim$ 50\,K (see Fig. \ref{fig2_RT} (c) inset), suggesting an anisotropic behavior of the $c$-axis and $b$-axis resistivity.

For all measured samples, La$_5$Co$_2$Ge$_3$ shows metallic behavior in the whole studied pressure range. For sample S2 measured in the PCC (see Fig. \ref{fig2_RT} (a)), the sharp drop of resistivity, associated with the FM transition, persists to pressures as high as 1.64\,GPa. The ferromagnetic transition temperature, $T_\textrm C$, is determined from the intersection of the two dashed lines as indicated in Fig.\,\ref{fig2_RT} (a). The dashed lines are drawn in the same way as described above (with multiple 1\,K windows on the high-temperature side over the temperature range of 5\,K - 10\,K to obtain the uncertainties). Using this criterion, we infer that, $T_\textrm C$ is suppressed from $\sim$ 4\,K to $\sim$ 3.4\,K upon increasing pressure from 0 to 1.64 GPa. At 1.97\,GPa, an anomaly with a different shape is observed at low temperatures. Upon cooling through $T\sim$ 6\,K, the resistivity shows a broad increase which is suggestive of superzone-gap formation. This feature implies that at 1.97\,GPa, La$_5$Co$_2$Ge$_3$ enters a low-temperature ground state below $T^*$ (defined below), which is different from the FM state at lower pressures. It appears likely that this new state is characterized by an antiferromagnetic component that partially gaps the Fermi surface\cite{Elliott1972,Friedel1987,Budko2000}. This superzone-gap-like feature in the resistivity is observed in all temperature-dependent data sets under pressures between 1.97\,GPa and 5.12\,GPa (see Fig. \ref{fig2_RT} (b) for data on sample S3 for $p\ge$ 2.74\,GPa taken in the mBAC with $j\parallel c$).

The transition temperature $T^*$, which is associated with the transition into this new state, is determined from the following construction of three lines in the low-, intermediate- and high-temperature regime as indicated in Figs. \ref{fig2_RT} (a) and (b). The low- and high-temperature lines are linear fits to the $\rho(T)$ data in these temperature regimes, whereas the intermediate-temperature line goes through the point of maximum slope of $\rho(T)$ and the slope corresponds to this maximum slope. $T^*$ is determined as the midpoint of the two intersection points of the dashed lines and the uncertainties of $T^*$ are obtained from the temperature difference of the two intersections points. Upon increasing pressure, $T^*$ first increases from $\sim$ 4.0\,K (2.74\,GPa) to $\sim$ 7.4\,K (4.10\,GPa), then decreases to $\sim$ 6.3\,K (4.73\,GPa) and finally increases again slightly to $\sim$ 6.4\,K (5.12\,GPa).

For sample S4, measured with $j\parallel b$, for all data sets under pressure up to 2.30\,GPa, resistivity decreases monotonically upon cooling from high temperatures, until it shows a sharp drop of resistivity when cooling through the phase transitions $T_\textrm C$ and $T^*$ (see Fig. \ref{fig2_RT} (c)). The corresponding transition temperature, $T_\textrm C$ ($T^*$), is determined from the intersection of the two dashed lines (drawn in the same way as described above) as indicated in Fig.\,\ref{fig2_RT} (c). This observation shows that the resistivity at the $T^*$ phase transition displays a distinct directional anisotropy, i.e., resistivity increases (decreases) upon cooling through $T^*$ along the $c$ ($b$) direction. The proposed superzone-gap formation outlined above is consistent with the $j\parallel c$ and $j\parallel b$ anisotropy of the resistive feature at $T^*$.

\begin{figure}
	\includegraphics[width=8.6cm]{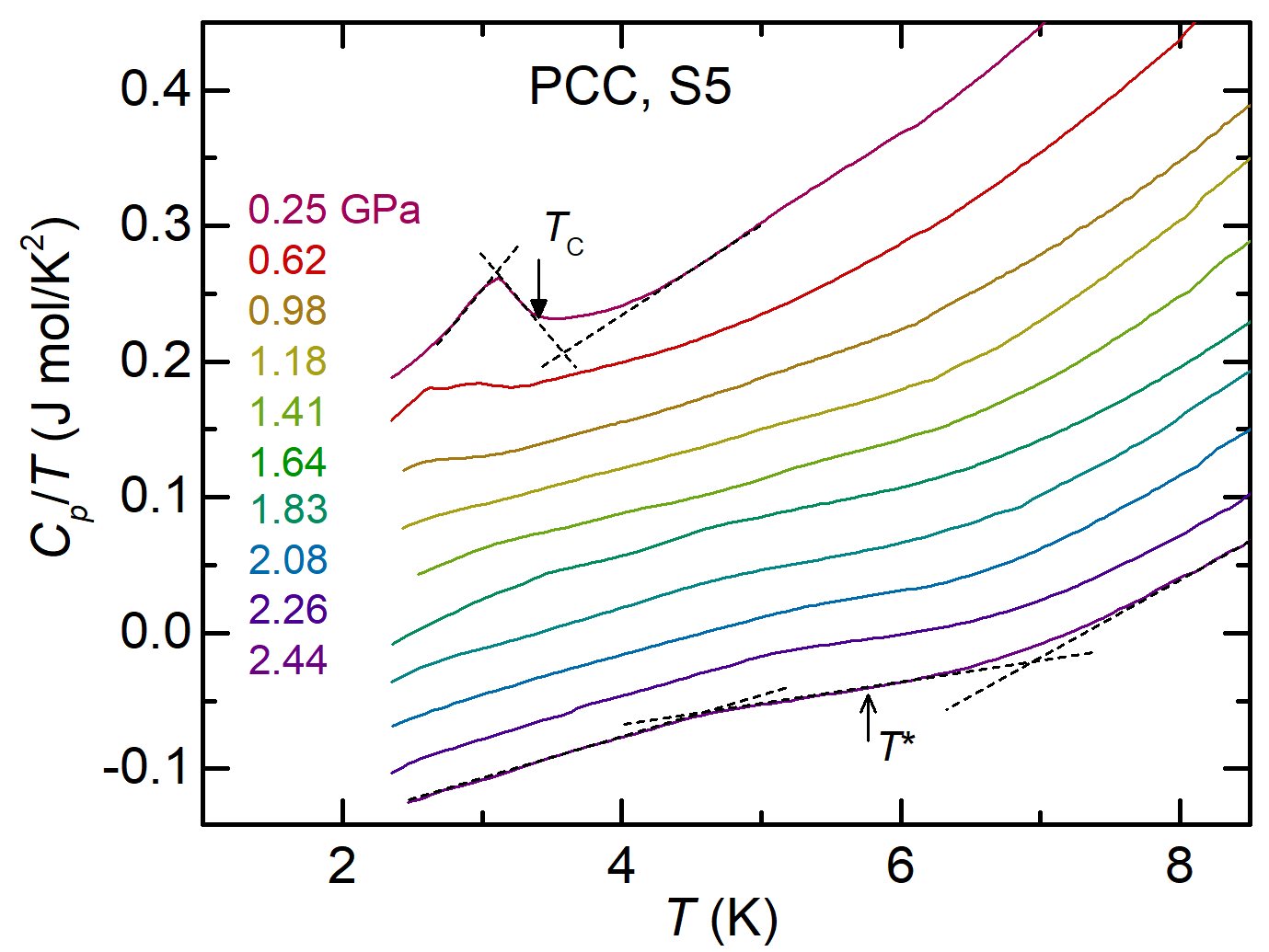}%
	\caption{Evolution of the temperature-dependent specific heat over temperature, $C_p$/$T$, of La$_5$Co$_2$Ge$_3$ with pressure up to 2.44 GPa in a piston-cylinder cell for sample S5. Criteria for the determination of the ferromagnetic transition temperature $T_\textrm C$ and the transition temperature into new ground state $T^*$ are indicated by dashed lines and arrows in the figures (see text for details). Data curves are shifted down by 0.03\,J mol/K$^2$ for clarity.
		\label{fig3_CT}}
\end{figure}

To further study the pressure effect on La$_5$Co$_2$Ge$_3$ from a thermodynamic perspective, specific heat measurements under pressure were performed. Figure \ref{fig3_CT} presents the specific heat divided by temperature, $C_p$/$T$, as a function of temperature for different pressures. At the lowest pressure measured (0.25\,GPa), a clear ``$\lambda$-shape" anomaly is observed at $\sim$ 3.3\,K, which is associated with the ferromagnetic transition. The shape of the anomaly is consistent with the second-order nature of the transition\cite{Saunders2020}. At 0.62\,GPa, the ``$\lambda$-shape" anomaly is suppressed to lower temperature at $\sim$ 3\,K and becomes significantly broader. In addition, a second feature at slightly lower temperature ($\sim$ 2.6\,K), the origin of which is unclear, is only observed for this pressure. We point out that in the resistivity measurements, shown in Fig. \ref{fig2_RT}, such a second feature at a similar pressure and temperature is not observed. We therefore did not include the second feature at 0.62\,GPa in the pressure-temperature phase diagram. At 0.98\,GPa, a single, broad anomaly is observed. For 1.18\,GPa $\le p \le$ 1.41\,GPa, $C_p$/$T$ displays a continuous, smooth change upon cooling. The reason for the absence of a clear thermodynamic feature in this pressure range despite the presence of clear resistive features, as presented above, is presently unknown. We speculate that in this pressure range, the change of entropy associated with the magnetic transition is broad in temperature and thus the specific heat feature is not resolvable from the non-magnetic background contribution. For $p\ge$ 1.64\,GPa, a broad hump-like feature is observed at $\sim$ 6\,K. Based on our previously-described observations in resistivity measurements, we associate this broad specific heat feature with the phase transition into the new type of order at high pressures. Thus, our thermodynamic, specific heat, measurements are consistent with the proposal that La$_5$Co$_2$Ge$_3$ enters a new state in the high-pressure, low-temperature region. The corresponding transition temperatures, $T_\textrm C$ and $T^*$, are determined from the intersections of the three dashed lines as indicated in Fig. \ref{fig3_CT} (constructed following the same way as the lines constructed above in resistivity measurements shown in Fig. \ref{fig2_RT} (b)).

\begin{figure}
	\includegraphics[width=8.6cm]{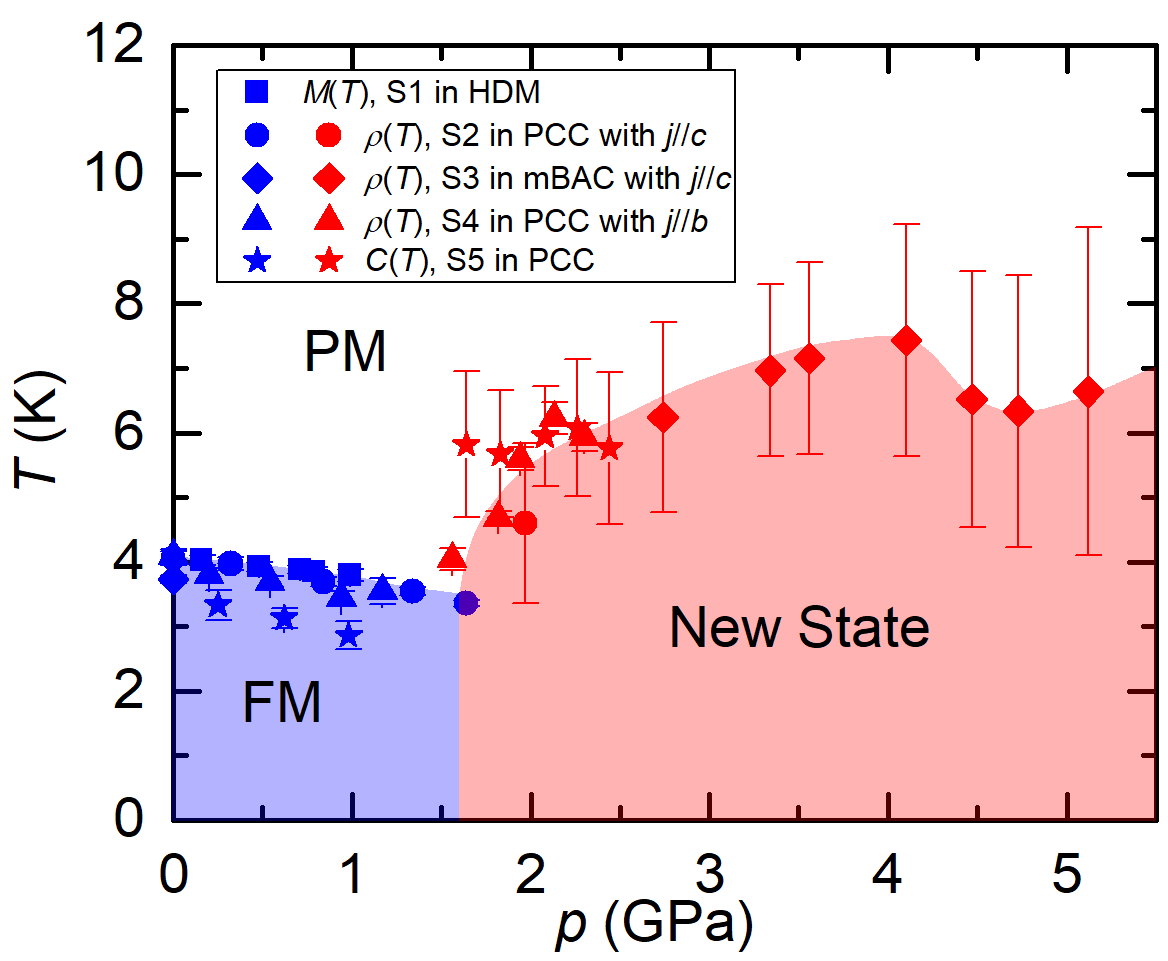}%
	\caption{Pressure-temperature ($p-T$) phase diagram of La$_5$Co$_2$Ge$_3$, as determined from magnetization (sample S1), resistivity (samples S2, S3, S4) as well as specific heat (sample S5) measurements. Transition temperatures $T_\textrm C$ (blue symbols) and $T^*$ (red symbols) are determined using the criteria shown in Figs.\,\ref{fig1_MT}-\ref{fig3_CT}. The determination of the error bars of the transition temperatures are described in detail in the text. The blue-shaded region corresponds to the region of ferromagnetic (FM) order, and the red-shaded region corresponds to the region of a new type of order. PM stands for paramagnetic.
		\label{fig4_pT}}
\end{figure}

The transition temperatures, $T_\textrm C$ and $T^*$, as determined from the magnetization, resistivity and specific heat measurements are used to construct a pressure-temperature ($p-T$) phase diagram, as shown in Fig. \ref{fig4_pT}. Overall, three phase regions exist in the studied $p-T$ phase space, and are separated by the determined phase transition lines $T_\textrm C(p)$ and $T^*(p)$. At high temperatures, La$_5$Co$_2$Ge$_3$ is in the paramagnetic (PM) state. In the low-temperature (below $T_\textrm C$) and low-pressure ($p\lesssim$ 1.7\,GPa) region, La$_5$Co$_2$Ge$_3$ is in the ferromagnetic state. The transition temperature $T_\textrm C$ is suppressed from $\sim$ 4.0\,K to $\sim$ 3.3\,K upon increasing pressure from 0\,GPa to $\sim$ 1.7\,GPa. In the low-temperature (below $T^*$) and high-pressure ($p\gtrsim$ 1.7 GPa) region, La$_5$Co$_2$Ge$_3$ shows a different type of order. The transition temperature $T^*$ manifests a nonmonotonic dependence on $p$ with a local maximum at $\sim$ 4.1\,GPa and a local minimum at $\sim$ 4.7\,GPa.

\begin{figure}
	\includegraphics[width=8.6cm]{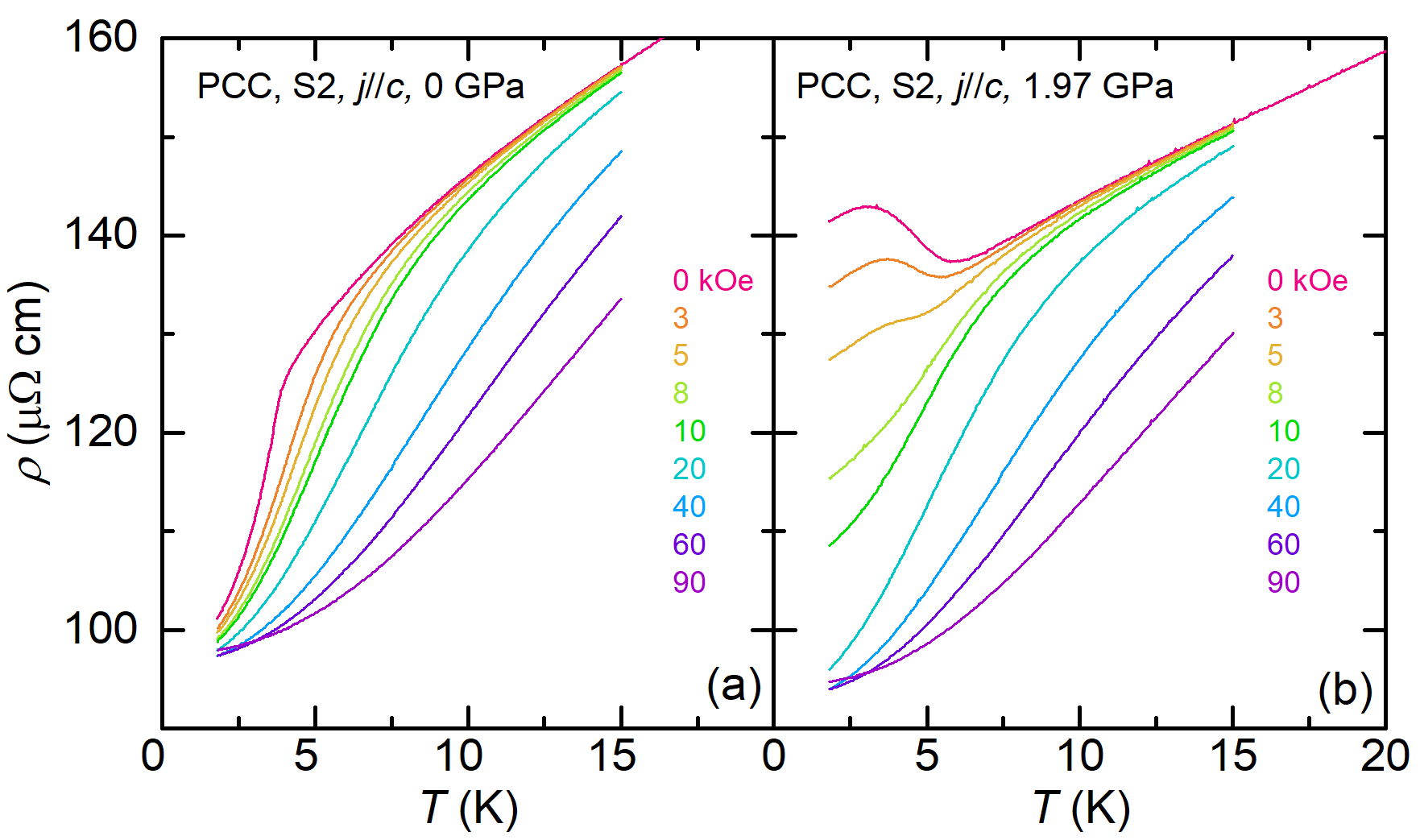}%
	\caption{Temperature-dependent resistivity of La$_5$Co$_2$Ge$_3$ in magnetic fields up to 90\,kOe (field was always applied perpendicular to the $bc$ plane) for sample S2 at 0\,GPa (a) and 1.97\,GPa (b). Current is applied along the crystallographic $c$ axis.
		\label{fig5_RTH}}
\end{figure}

To further investigate the nature of the new type of order at high pressures and low temperatures, we studied the response of the superzone-gap feature to external magnetic fields. Figure \ref{fig5_RTH} presents the temperature-dependent resistivity, $\rho(T)$, in magnetic fields up to 90\,kOe, applied perpendicular the $bc$ plane, for sample S2 at 0\,GPa and 1.97\,GPa. At low pressures, when magnetic field is increased, the resistive anomaly broadens and shifts to higher temperature. This is consistent with the expectation when the external magnetic field is applied along the ferromagnetic easy axis\cite{Saunders2020}. At high fields, the $\rho(T)$ behavior is consistent with La$_5$Co$_2$Ge$_3$ undergoing a crossover to a fully spin-polarized state upon cooling. At high pressures, where our data demonstrate a phase transition into a state with different type of order, the resistive anomaly is broadened with applying magnetic field but the apparent transition temperature does not shift very much for low fields. At high fields, the resistivity displays a similar temperature dependence compared to that at low pressures and under high magnetic fields. The data in Fig. \ref{fig5_RTH} (b), then, are consistent with a low-field antiferromagnetic state that becomes a high-field spin-polarized state when the external field is applied along the antiferromagnetic hard axis.

\section{Conclusion}
In summary, magnetization, resistivity and specific heat measurements under pressure up to 5.12\,GPa were performed on single-crystalline La$_5$Co$_2$Ge$_3$. The ambient-pressure ferromagnetic transition temperature, $T_\textrm C$, is suppressed upon increasing pressure up to $\sim$1.7\,GPa. Instead of $T_\textrm C$ being suppressed further upon increasing pressure beyond 1.7\,GPa, we find that La$_5$Co$_2$Ge$_3$ enters a different low-temperature ground state. The transition temperature, $T^*$, into the new state has a non-monotonic dependence on $p$ up to 5.12 GPa. Overall, our study shows that La$_5$Co$_2$Ge$_3$ manifests another example of avoided ferromagnetic quantum criticality in a metallic system via the appearance of a new ordered state. Based on our transport data in zero and finite field, it seems likely that this new type of order is magnetic in nature with an antiferromagnetic component. To clarify the exact nature of the new phase, microscopic studies, such as neutron scattering or $\mu$SR under pressure, would be needed.


\begin{acknowledgements}
This work was supported by the US Department of Energy, Office of Science, Basic Energy Sciences, Materials Sciences and Engineering Division. Ames Laboratory is operated for the US Department of Energy by Iowa State University under Contract No. DE-AC02-07CH11358. L. X. and E. G. were supported, in part, by the Gordon and Betty Moore Foundation’s EPiQS Initiative through Grant No. GBMF4411. L. X. was also supported, in part, by the W. M. Keck Foundation.
\end{acknowledgements}

\section{Appendix}



\begin{figure}
	\includegraphics[width=8.6cm]{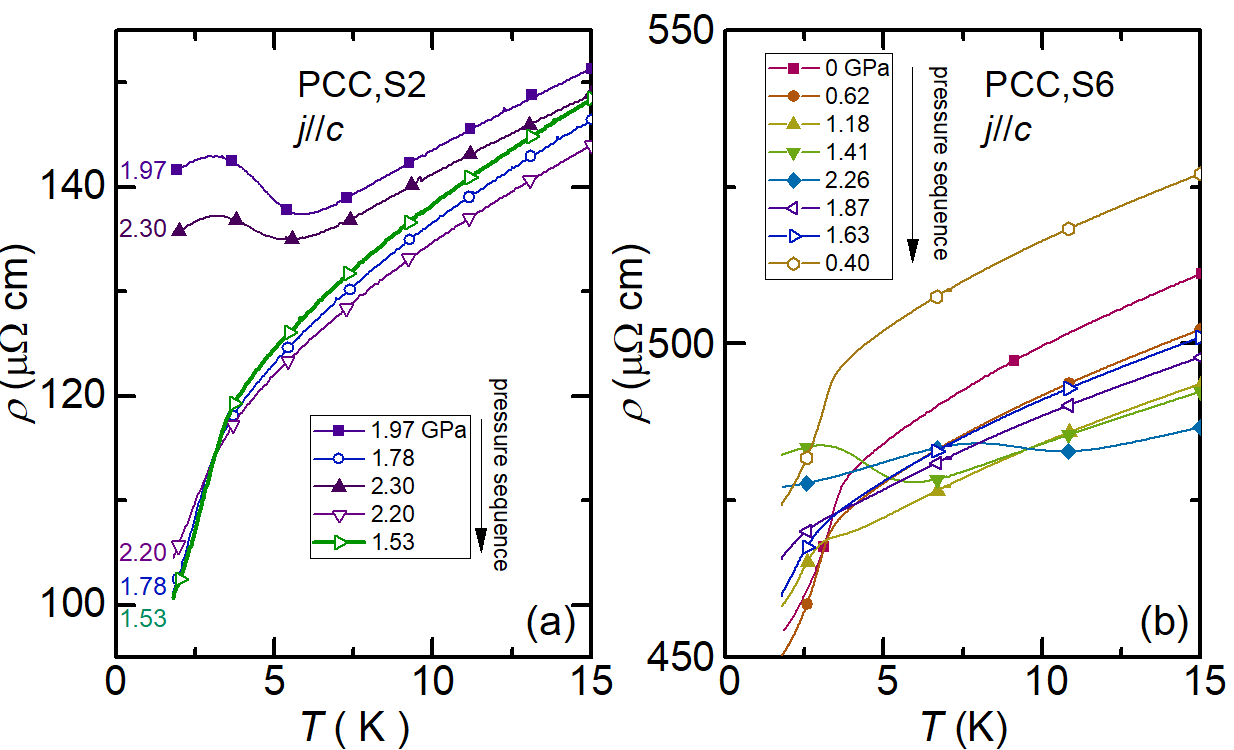}%
	\caption{Temperature-dependent resistivity for samples S2 (a) and S6 (b) of La$_5$Co$_2$Ge$_3$ measured with current applied along $c$ direction in piston-cylinder cell where pressure is changed in a non-monotonic way. Solid (open) symbols correponds to pressure increase (decrease) from a previous measurement. The corresponding pressure change sequences are indicated by arrows in the figures.
		\label{S1_RT}}
\end{figure}

In the following, we present results of further resistivity measurements on La$_5$Co$_2$Ge$_3$ under increasing and decreasing pressure. These measurements indicate that whereas La$_5$Co$_2$Ge$_3$ enters into a new state in the high-pressure, low-temperature region, the exact critical pressure, which separates the FM and the new ground state, as well as transition temperature, $T^*$, can vary somewhat from sample to sample and depends on the history of pressure change.

In the main text, Fig. \ref{fig2_RT} (a) shows the $\rho(T)$ for sample S2 measured in the PCC with $j\parallel c$ where pressure is monotonically increased to 1.97 GPa. Further measurements on this sample were performed where pressure was changed non-monotonically after 1.97 GPa and the results are shown in Fig. \ref{S1_RT} (a). We start our discussion at 1.97 GPa, where we find clear evidence for the superzone-gap-like feature in resistivity, and now turn to the next pressure point, which was obtained by decreasing pressure to 1.78 GPa. This results, as expected, in a phase transition back into the FM state at low temperatures. Increasing pressure again to 2.30\,GPa leads, again, to the observation of the superzone-gap-like feature. Then, surprisingly, when reducing the pressure back to 2.20\,GPa, we observe a resistive behavior which we would associate with the low-pressure behavior of FM ordering instead of the superzone-gap-like feature. We would have not expected this result based from our phase diagram. These data suggest that the pressure history seems to affect the critical pressure.


To investigate the dependence of the critical pressure in a more systematic way, sample S6 was measured in the PCC with $j\parallel c$, where pressure is first monotonically increased and then monotonically decreased. The $\rho(T)$ data for selected pressures are presented in Fig. \ref{S1_RT} (b). We point out that S6 has a higher residual resistivity, $\rho_0$, compared with other measured samples, indicating a somewhat higher level of disorder in this sample. At low pressures, $\rho(T)$ displays a sharp drop upon cooling, which corresponds to the FM transition. With increasing pressure to 1.41 GPa and higher, a clear increase of $\rho$ upon cooling is observed, suggesting that La$_5$Co$_2$Ge$_3$ enters into the new ordered state. When pressure is monotonically decreased from the highest pressure, we see that at 1.63 GPa, the superzone-gap-like feature is lost and a sharp drop of resistive anomaly, which we associate with the FM transition, is observed. Upon further decreasing pressure, sample S6 stays FM at low temperature. These measurement results demonstrate that the critical pressure upon increasing and decreasing pressure are clearly different for S6 ($\sim$ 1.41 GPa and $\sim$ 1.63 GPa with increasing and decreasing pressure). We further point out that even upon increasing pressure, the critical pressure for S6 ($\sim$ 1.41 GPa) is lower than for S2 ($\sim$ 1.7 GPa).

\begin{figure}
	\includegraphics[width=8.6cm]{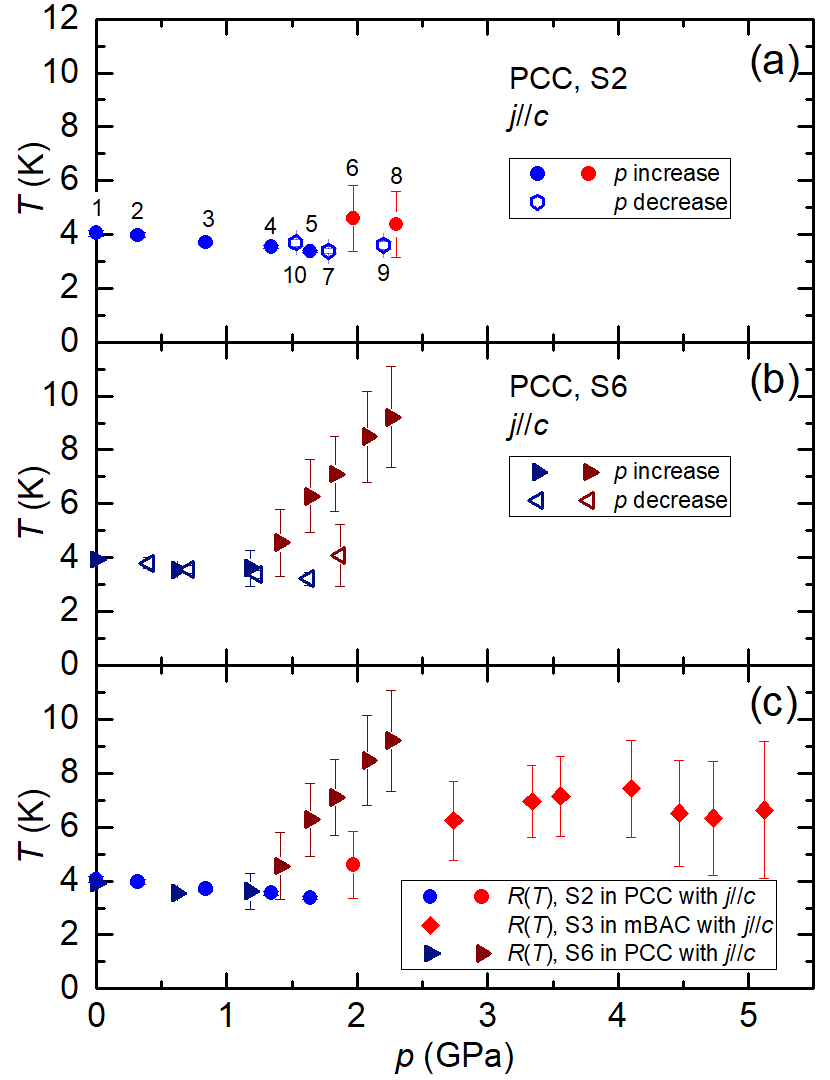}%
	\caption{Pressure-temperature ($p-T$) phase diagrams of La$_5$Co$_2$Ge$_3$ determined from measurements on sample S2 (a), sample S6 (b), samples S2, S3 and S6 (c). Solid (open) symbols correspond to data that were obtained after increasing (decreasing) pressure with respect to the previous measurement. Numbers in (a) indicate the sequence of pressure change.
		\label{S2_phase_diagram}}
\end{figure}


The corresponding transition temperatures $T_\textrm C$ and $T^*$, determined from the measurements on samples S2 and S6 are summarized in Figs. \ref{S2_phase_diagram} (a) and (b), respectively. The transition temperatures determined from resistivity measurements with $j\parallel c$ (samples S2, S3 and S6), where pressure is monotonically increased, are plotted in Fig. \ref{S2_phase_diagram} (c) together for comparison. Whereas the pressure dependence of the FM transition temperature, $T_\textrm C$, agrees well with each other for all different samples and experiments, the critical pressure varies from sample to sample and depends on the history of pressure change. In addition, the corresponding transition temperature, $T^*$, also varies ($T^*$ is $\sim$ 4.6\,K and $\sim$ 7.8\,K for S2 and S6 respectively, at a pressure of $\sim$ 2\,GPa). Overall, whereas the basic features of the $p-T$ phase diagram of La$_5$Co$_2$Ge$_3$ are robust among all measurements (i.e., La$_5$Co$_2$Ge$_3$ is ferromagnetic in the low-temperature, low-pressure region and enters into a new state in the low-temperature, high-pressure region), the sensitivity of the pressure-induced transition to the super-zone-gapped state to the pressure history as well as possibly small differences in degrees of disorder suggests that there are parameters influencing the precise values of the critical pressure as well as $T^*$ that still need to be understood.

\clearpage

\bibliographystyle{apsrev4-1}
%

\end{document}